\documentclass[aps,prd]{revtex4}
\bibliography{apsrev}
\begin{document}

	\title{Conformally Invariant Gauge Theory of $3$-Branes in $6$-$D$
	       and the Cosmological Constant}
\author{E.I.Guendelman}
\email{guendel@bgumail.bgu.ac.il}
\affiliation{Physics Department, Ben Gurion University, Beer Sheva,
	Israel}

\author{E. Spallucci}
\email{spallucci@trieste.infn.it}
\affiliation{Dipartimento di Fisica Teorica, Universit\`a di Trieste,
		          and INFN, Sezione di Trieste}

\date{\today}
\begin{abstract}
	It is shown that the gauge theory of relativistic $3$-Branes can be
	formulated in a conformally invariant way if the embedding space is
	six-dimensional. The implementation of conformal invariance requires the
	use of a modified measure, independent of the metric in the action.
	Brane-world scenarios without the need of a cosmological constant in $6D$
	are constructed. Thus, no ``old'' cosmological constant problem 
	appears at this level. 
	\end{abstract}
 \maketitle

	\section{Introduction}
	Extended objects of various dimensions are present in the modern
	formulation of string theory. Among the various kind of branes
	a unique role is played by $D$--branes \cite{dbrane} as they
	can trap the end-points of open strings. $D$-brane inspired cosmological
	models, commonly termed ``brane--universe'' models, are currently
	under investigation as they seem to offer a possible solution to the
	longstanding hierarchy problem in gauge theories.\\
	We think that the
	gauge theory formulation of $p$--branes, proposed some years ago
	as an alternative to the standard description
	of relativistic extended objects \cite{aae},\cite{gtm}, is well suited
	to describe this new type of cosmological scenario. Furthermore,
	the description of $p$--branes in terms of associated gauge potentials
	offers a vantage point to study some specific problem 
	as the one concerning the fine tuning of the cosmological constant.\\
	Here we will see that for $3$-branes considered in an embedding $6D$
	space the gauge theory formulation of $3$-branes allows a conformally
	invariant realization. An essential element necessary to implement
	conformal invariance is the introduction of a measure of integration in
	the action which is independent of the metric
	\cite{g3},\cite{g4},\cite{g5}, \cite{g6}. 
	We use then such a formulation to construct a new type of brane world
	scenario.\\
	Brane world scenarios in general are concerned with the possibility that
	our universe is built out of one or more $3$-branes living in some
	higher dimensional space, plus some bulk component \cite{7},
	\cite{8}, \cite{9}, \cite{10}, \cite{11}.
	In particular, the possibility of $3$-branes embedded in $6D$ space has
	been studied in \cite{12},  \cite{13}, \cite{14}, \cite{15}, \cite{16}.
	In this case the effect of the tension of the branes is to induce
	curvature only in the extra dimensions. In these models there is still a
	question of fine tuning that has to be addressed, since although the
	branes themselves do not curve the observed four dimensions, the bulk
	components of matter do, and they have to be fine tuned in order to get
	(almost)zero four dimensional vacuum energy. This very special feature
	of $3$-branes is a $6D$ embedding spacetime is related to the fact that
	such matter content, even coupled to gravity, has a conformal invariance
	associated to it.\\
	The strategy that we will follow in order to solve this problem is to
	incorporate the ``brane-like features'' that are quite good in what
	concerns the cosmological constant problem into the ``bulk'' part of the
	brane scenario as well. In this way both bulk and singular brane
	contributions will share the fundamental feature of curving only the
	extra dimensions. The gauge field formulation of $3$-branes in $6D$ 
	(GFF3B6D) is ideally suited for such a program.\\
	As we will see GFF3B6D allows us to understand , extend and give a 
	``pure brane interpretation'' of the results of \cite{17}, where a
	``square root gauge theory'', coupled to $3$-branes in $6D$ was
	considered. This model has conformal invariance and there is no need
	to introduce a $6D$ cosmological constant. The ``fundamental physics''
	behind the model is not so clear however and its different matter
	elements: gauge fields, $3$-branes appear rather disconnected from each
	other. We will see that GFF3B6D allows us resolve these drawbacks and
	present a more general set of brane-world solutions, where the solutions
	presented in \cite{17} appear as very particular cases.
	The paper is organized as follows. In Sect.II, we go through the gauge
	formulation of branes; in Sect.III, we study the GFFB6D and show that
	this system displays conformal invariance when a modified measure is
	introduced, also the dual picture to this formulation is introduced; 
	in Sect.IV, the equations of motion in this dual picture are studied. We
	end up with a brief discussion  and conclusions.

	\section{Summary of gauge formulation of branes}
	In this section we would like to recall some features of the
	``gauge formulation'' of brane dynamics introduced some years ago
	by one of the authors (E.S.). A possible way to bridge the gap between
	extended objects and gauge field is to provide a non-geometric
	 description of the former ones in terms of appropriate field variables.
	This model has been investigated in depth in \cite{aae}, \cite{gtm}.
	Here we do not report about what has been already published but add
	some new matter related to $D$-branes. \\
	The effective description of extended objects in terms higher rank
	gauge forms requires the introduction of both a \textit{slope field}
	and a \textit{geodesic field}. The slope field defines the tangent
	(hyper)plane to brane world surface in target spacetime. The geodesic
	field represents the brane field strength. As such it solves a set
	of Bianchi identities which reproduce the classical equation of
	motion once are projected on the brane world surface with the aid
	of the geodesic field. \\
	
	\subsection{ Brane slope field and gauge potential.}
	
	A $p$-brane can be described as $p+1$ dimensional hypersurface embedded
	into an higher $D$-dimensional target spacetime. The parametric equations
	describing the brane world-volume reads
	$x^M=Y^M\left(\, \sigma^0\ , \vec \sigma\,\right)$, where $\vec \sigma
	=\left(\, \sigma^0\ , \sigma^1\ ,\dots\ ,\sigma^p\right)$, with
	$p+1\le 1+d\equiv D $ and  $x^M=\left(\, x^0\ , x^1\ ,\dots \ ,\right)$ are
	target spacetime coordinates. To each point of the brane world-volume
	one can define a tangent hyperplane. The orientation of the tangent
	hyperplane is encoded into the current density
	
	\begin{equation}
	 J^{{}^{M_1\dots M_{p+1}}}(x) =
	\int d^{p+1}\sigma\, \delta^{(d+1)}\left[\, x-Y(\sigma)\,\right]
	\epsilon^{m_1\, m_2\, \dots\, m_{p+1}}\partial_{m_1}Y^{{}^{M_1}}\dots
	\partial_{m_{p+1}}Y^{{}^{M_{p+1}}}
	\end{equation}
	The Dirac-delta makes $J$ to be ``singular'' , i.e. $J$
	different from zero only along the brane world-volume.
	However, we can define a ``\textit{slope field}'' 
	$F^{{}^{M_1\dots M_{p+1}}}(x)$   as a ``regular'' field
	whose restriction on the brane world-volume  matches the
	the tangent hyperplane:
	
	\begin{equation}
	\int d^{d+1}x\, \delta^{(d+1)}\left[\, x-Y(\sigma)\,\right]
	F^{{}^{M_1\dots M_{p+1}}}\left(\, x\,\right)=
	\epsilon^{m_1\, m_2\, \dots\, m_{p+1}}\partial_{m_1}Y^{{}^{M_1}}
	\dots \partial_{m_{p+1}}Y^{{}^{M_{p+1}}}
	\end{equation}
	
	Accordingly, the current density can be written as
	
	\begin{equation}
	 J^{{}^{M_1\dots M_{p+1}}}(x) =
	\int d^{p+1}\sigma\, \delta^{(d+1)}\left[\, x-Y(\sigma)\,\right]\,
	F^{{}^{M_1\dots M_{p+1}}}\left(\, x\,\right)
	\end{equation}
	
	A gauge-type formulation of brane dynamics can be recovered from
	the action 
	
	\begin{eqnarray}
	S=&& e^2\int d^{d+1}x\,\sqrt{-g_{(d+1)}}
	\left[\, -\frac{1}{(p+1)!}\, g_{{}_{M_1\, N_1} }\dots 
	g_{{}_{M_{p+1}\, N_{p+1}}}\,
	W^{{}^{M_1\dots M_{p+1}}} 
	W^{{}^{N_1\dots N_{p+1}}}\,\right]^{1/2}\nonumber\\
	 && +\frac{1}{(p+1)!}\int d^{d+1}x\,
	\sqrt{-g_{(d+1)}}\, W^{{}^{M_1\dots M_{p+1}}} 
	\partial_{{}_{[\, M_1}}\, B_{{}_{M_2\dots M_{p+1\,]}}} \label{sact}
	\end{eqnarray}

	where $W$ is  a contravariant, totally antisymmetric tensor, and
	the curl of the gauge potential $B$ is the geodesic field we
	mentioned above. A target spacetime metric $g_{{}_{M_1\, M_2}}$ 
	has been introduced to make the model generally covariant.
	$e^2$ is a  constant with mass-squared dimension
	providing to $W$ the canonical dimension of a gauge field strength
	in a $d+1$ dimensional spacetime.
	The $B$-field plays the role of Lagrange multiplier
	which imposes $W$ to be divergence-free:
	
	\begin{equation}
	\frac{\delta S}{\delta B_{ {}_{M_2\dots M_{p+1}}}} =0\longrightarrow
	\partial_M \, \left(\, \sqrt{-g_{(d+1)}}\,
	W^{{}^{M_1\dots M_{p+1}}} \,\right) =0 \label{div}	
	\end{equation}
	
	The divergence in (\ref{div}) is a generally covariant	operator
	built up with the aid of an appropriate affine connection.
	The field equation (\ref{div}) has a general solution of the form
	
	\begin{equation}
	W^{{}^{M_1\dots M_{p+1}}}=\frac{1}{(D-p-1)!}
	\frac{1}{\sqrt{-g_{(d+1)}}}
	\epsilon^{{}^{M\, M_2\dots M_{p+1}\,M_{p+2}}\dots
	 M_D }\partial_{{}_{[\, M_{p+2}}}\, A_{ {}_{M_{p+3}\dots M_{D\,]}}} +
	\frac{e^{p-1}}{\sqrt{-g_{(d+1)}}} \,J^{{}^{M\, M_2\dots M_{p+1}}}
	\label{div0}
	\end{equation}
	
	The ``regular'' part of $W$ has been written in terms of
	a $D-p-1$ gauge potential, while the ``singular'' part is expressed
	in terms of the current density. The constant $e^{p-1}$ provides
	the matching between the (different) canonical
	dimensions of $W$ and $J$. In order to satisfy equation (\ref{div0})
	$J$ must be divergence-free.
	The condition $ \partial_M\, J^{{}^{M\, M_2\dots M_{p+1}}}=0$ selects
	boundary-free branes, i.e. closed or infinitely extended objects.
	Having in mind a cosmological scenario where branes are orthogonal
	to the extra dimensions,  we specialize our model in the following
	way:\\
	i) $Y^\mu(\sigma)$ are the embedding functions of a $D_p$--brane, 
	i.e. a static topological defect in the spacetime fabric located at 
	a definite position;\\
	ii) only gravity is free to propagate-off the brane. Thus, the regular
	part of the $W$-field must be zero, the ``singular'' part of $W$
	being strictly localized on the brane. Hence, $W=J$.\\ 
	
	\subsection{Relation with the world-volume dynamics}
	
	After the brief discussion of  the gauge formulation
	of $p$-brane, summarized above,  
	 it may be worth to provide the relations to the more familiar
	 world volume description of $p$-brane dynamics. \\
	 Let us compute the classical action for the solution (\ref{div0}).
	 
	\begin{eqnarray}
	S\left[\, J\,\right]=&& e^2\int d^{d+1}x\,\sqrt{-g_{(d+1)}}
	\left[\, -\frac{1}{(p+1)! \left(\,\sqrt{-g_{(d+1)}}\,\right)^2 }\, 
	e^{2p-2}\, g_{{}_{M_1\, N_1}}\dots g_{{}_{M_{p+1}\, N_{p+1}}} 
	J^{{}^{M_1\dots M_{p+1}}} 
	J^{{}^{N_1\dots N_{p+1}}}\,\right]^{1/2}\nonumber\\
	=&& e^{p+1}\int d^{d+1}x\,
	\left[-\frac{1}{(p+1)!}\, 
	\int d^{p+1}\sigma\, \delta^{(d+1)}\left[\, x-Y(\sigma)\,\right]
	\epsilon^{m_1\, \dots\, m_{p+1}}\partial_{m_1}\, Y^{{}^{M_1}}\dots
	\partial_{m_{p+1}}\, Y^{{}^{M_{p+1}}}\, \right. \times\nonumber\\
	&&
	g_{{}_{M_1\, N_1}}\dots g_{{}_{M_{p+1}\, N_{p+1}}}\,
	\int d^{p+1}\tilde\sigma\, \delta^{(d+1)}
	\left[\, x-Y(\tilde\sigma)\,\right]
	 \epsilon^{n_1\,  \dots\, n_{p+1}}\,\left. \partial_{n_1}\, 
	 \,Y^{{}^{N_1}}\dots \partial_{n_{p+1}}\, Y^{{}^{N_{p+1}}} 
	\, \right]^{1/2}\nonumber\\
	=&& e^{p+1}\int d^{d+1}x\,
	\sqrt{
	-\frac{1}{(p+1)!}\, F^{{}^{M_1\dots M_{p+1}}}\,
	                    F_{{}^{M_1\dots M_{p+1}}}
	     }\times\nonumber\\
	&& \left[\, 
	\int d^{p+1}\sigma\, \delta^{(d+1)}\, \left[\, x-Y(\sigma )\,\right] 
	\int d^{p+1}\tilde\sigma\, \delta^{(d+1)}\left[\, x-Y(\tilde\sigma )\,
	\right]\,\right]^{1/2}
	\end{eqnarray}
	
	Invariance under reparametrization of the brane world volume allows
	us to write
	
	\begin{eqnarray}
	\left[\,
	\int d^{p+1}\sigma\, \delta^{(d+1)}\left[\, x-Y(\sigma)\,\right] 
	\int d^{p+1}\tilde\sigma\, \delta^{(d+1)}
	\left[\,x-Y(\tilde\sigma)\,\right]
	\,\right]^{1/2}
	&&= \left[\,\left(\,\int d^{p+1}\sigma\, \delta^{(d+1)}\left[\,
	x-Y(\sigma)\,\right]\,\right)^2\,\right]^{1/2}\nonumber\\
	&&=\int d^{p+1}\sigma\, \delta^{(d+1)}\left[\,
	x-Y(\sigma)\,\right]
	\end{eqnarray}
	
	Thus,
	
	\begin{eqnarray}
	S\left[\, J\,\right]=&& e^{p+1}\int d^{d+1}x\,
	\sqrt{-\frac{1}{(p+1)!}\, F^{{}^{M_1\dots M_{p+1}}}\,
	F_{{}^{M_1\dots M_{p+1}}\,}} 
	\int d^{p+1}\sigma\, \delta^{(d+1)}\left[\,
	x-Y(\sigma)\,\right]\nonumber\\
	=&& e^{p+1}\, \int d^{d+1}x\,\int d^{p+1}\sigma\, 
	\delta^{(d+1)}\left[\,x-Y(\sigma)\,\right]
	\sqrt{-\frac{1}{(p+1)!}\, F^{{}^{M_1\dots M_{p+1}}}\,
	F_{{}^{M_1\dots M_{p+1}}\,}}
	\nonumber\\
	=&& e^{p+1}\, \int d^{p+1}\sigma\, 
	\left[ -\frac{1}{(p+1)!}\, 
	\epsilon^{m_1\, \dots\, m_{p+1}}\partial_{m_1}Y^{{}^{M_1}}\dots
	\partial_{m_{p+1}}Y^{{}^{M_{p+1}}}
	\epsilon^{n_1\,, \dots\,
	n_{p+1}}\partial_{n_1}Y^{{}^{N_1}}\dots
	\partial_{n_{p+1}}Y^{{}^{N_{p+1}}}\,\right]^{1/2}
	\label{sclass }
	\end{eqnarray}
	
	By introducing the induced metric $\gamma_{mn}$ as
	
	\begin{equation}
	\gamma_{mn}\left(\,\sigma\,\right)\equiv g_{MN}\, \partial_m\, Y^M\, 
	\partial_n\, Y^N
	\end{equation}
	
	we can write the argument of the square root in (\ref{sclass }) in a
	more transparent form. 
	
	\begin{eqnarray}
	\epsilon^{m_1\, \dots\, m_{p+1}}\partial_{m_1}Y^{{}^{M_1}}\dots
	\partial_{m_{p+1}}Y^{{}^{M_{p+1}}}\epsilon^{n_1\,  \dots\,
	n_{p+1}}\partial_{n_1}Y^{{}^{N_1}}\dots
	\partial_{n_{p+1}}Y^{{}^{N_{p+1}}}&&=
	\epsilon^{m_1\, \dots\, m_{p+1}}\epsilon^{n_1\, \dots\,
	n_{p+1}}\, \gamma_{m_1\, n_1}\, \dots \, \gamma_{m_{p+1}\, n_{p+1}}
	\nonumber\\
	&&=(p+1)!\mathrm{Det}\left(\, \gamma_{mn}\,\right)
	\end{eqnarray}
	
	Accordingly, the classical action turns out to be just the standard
	world volume action
	
	\begin{equation}
	S= e^{p+1}\int d^{p+1}\sigma\,\sqrt{-\mathrm{Det}\left(\,
	\gamma_{mn}\,\right) } \label{NG}
	\end{equation}
	
	where $e^{p+1}$ plays the role of brane tension.\\
	The induced dynamics of a $D_p$-branes is encoded into a more
	complex effective action, i.e.  a Born-Infeld type functional \cite{BI}

	\begin{equation}
	S= \mathrm{Tension}\int d^{p+1}\sigma\,\sqrt{-\mathrm{Det}\left(\,
	\gamma_{mn} + \frac{1}{e^2} F_{mn}\,\right) }\label{BI}
	\end{equation}
	
	where, $F_{mn}=\partial_{[\, m}\, A_{ n\,] }$ is the field strength of 
	a $U\left(\, 1\,\right)$ gauge field. Dirichlet branes fit 
	in our gauge-type description as well. 
	We notice that the Born-Infeld action (\ref{BI})
	can be recovered in the same way we obtained (\ref{NG}) provided
	one replaces $g_{{}_{M_1\, N_1}}$ in (\ref{sact})  according
	with the rule
	
	\begin{equation}
	g_{{}_{M_1\, N_1}}\longrightarrow g_{{}_{M_1\, N_1}} +\frac{1}{e^2}
	F_{{}_{M_1\, N_1}}
	\end{equation}
	
	where, $F_{{}_{M_1\, N_1}}$ is a target spacetime
	field related to $F_{mn}$ by
	
	\begin{equation}
	F_{mn}=F_{{}_{M_1\, N_1}}\, \partial_m\, Y^{{}^{M_1}}\, \partial_n\,
	Y^{{}^{N_1}}\, 
	\end{equation}
	
	We shall not elaborate anymore on this result which is not essential
	for what follows. In the next sections we shall drop out the term
	$F/e^2$ and focus on the gravitational effects only.\\
	Finally, we recall that the special case $p=3$ can be made Weyl
	invariant through the introduction of the \textit{dilaton} field 
	\cite{weyldp}. This field acts as a St\"uckelberg 
	conpensator \cite{stuck}
	under local rescaling of various fields . In the next
	section we shall describe an alternative, \textit{dynamical} mechanism,
	to make the $D_3$-brane action Weyl invariant. Instead of a 
	\textit{fundamental} scalar degree of freedom we shall consider
	a ``composite'' scalar density replacing the gravitational volume
	term $\sqrt{-g_{(5+1)}}$.

	\subsection{$D$-brane gauge formulation}
	The parametric equations for a $D_p$--brane are usually
	written by splitting the coordinates into ``parallel'' , 
        and ``transverse'' :
        
        \begin{eqnarray}
	&& x^M=\left(\, x^\mu_{||}\ , x^k_\perp\,\right)\\
	&& x^\mu_{||}=Y^\mu_{||}\left(\, \sigma^0\ , \vec \sigma\,\right)\\
	&& x^k_\perp=Y^k_\perp\left(\,\vec \sigma\,\right)\\
	&& 0 \le \mu\le p+1\ ,\quad p+2\le k\le D
	\end{eqnarray}
	
	As a first example, let us consider a flat  ``slab'' at the origin
	of transverse space, that is we choose
	
	\begin{equation}
	Y^k_\perp\left(\,\vec \sigma\,\right)=0
	\end{equation}
	
	and $g^{(3+1)}_{\mu\nu}=\eta_{\mu\nu}$.
	The $D_p$--brane current density is of the form
	
	\begin{eqnarray}
	 J^{ {}^{M_1\dots M_{p+1}}}&&=\int d^{p+1}\sigma\,
	  \delta^{(p+1)}\left[\,
	 x^\mu_{||} - Y^\mu_{||}\left(\, \sigma^0\ , \vec \sigma\,\right)\,
	 \right]\,
	 \delta^{(d-p)}\left(\, \vec x_\perp\,\right)\times\nonumber\\
	&& \delta^{{}^{M_1}}_{\mu_1}\dots\delta^{{}^{M_{p+1}}}_{\mu_{p+1}}
	 \epsilon^{m_1\dots m_{p+1}}\partial_{m_1} Y^{\mu_1}_{||}\dots
	 \partial_{m_{p+1}} Y^{\mu_{p+1}}_{||}\nonumber\\
	&&= \delta^{{}^{M_1}}_{\mu_1}\dots\delta^{{}^{M_{p+1}}}_{\mu_{p+1}}
	 \delta^{(d-p)}\left(\, \vec x_\perp\,\right)
	\int dY^{\mu_1}_{||}\wedge\dots\wedge dY^{\mu_{p+1}}_{||}
	\delta^{(p+1)}\left[\,
	 x^\mu_{||} - Y^\mu_{||}\left(\, \sigma^0\ , \vec \sigma\,\right)\,
	 \right]\nonumber\\
	&&=\delta^{{}^{M_1}}_{\mu_1}\dots\delta^{{}^{M_{p+1}}}_{\mu_{p+1}}
	\delta^{(d-p)}\left(\, \vec x_\perp\,\right)\, J^{\mu_1\dots\mu_{p+1}}
	\left(\, x_{||}\,\right)
	\end{eqnarray}
	
	It is immediate to generalize  the current to the case where a second
	$D_p$--brane is present in different point, say $\vec x_\perp = \vec
	Y_\perp $. Then,

	\begin{equation}
	J^{M_1\dots
	M_{p+1}}=\delta^{{}^{M_1}}_{\mu_1}\dots\delta^{M_{p+1}}_{\mu_{p+1}}
	\left[\, \delta^{(d-p)}\left(\, \vec x_\perp\,\right)+
	\delta^{(d-p)}\left(\, \vec x_\perp - \vec Y_\perp\,\right)\,\right]
	J^{\mu_1\dots\mu_{p+1}}\left(\, x_{||}\,\right)
	\end{equation}
	
	By iterating this procedure we can add as many $D$-branes we
	want and even to consider a continuous distribution of them.
	In the latter case
	
	\begin{equation}
	J^{{}^{M_1}\dots
	{}^{M_{p+1}}}=\delta_{\mu_1}\dots\delta^{{}^{M_{p+1}}}_{\mu_{p+1}}
	\rho\left(\, \vec x_\perp\,\right)
	J^{\mu_1\dots\mu_{p+1}}\left(\, x_{||}\,\right)
	\label{bdens}
	\end{equation}
	
	where $\rho\left(\, \vec x_\perp\,\right)$ represents the
	density of $D_p$--branes in the transverse space.\\
	In the original papers \cite{aae}, \cite{gtm} gravity was introduced  by
	making the action (\ref{sact}) generally covariant and by adding
	an Einstein term. This procedure will be generalized in the
	next section in such a way to endow the model with conformal 
	invariance, possible in the case of $3$-branes embedded in $6D$
	spacetime.

	\section{Conformally Invariant Realization in $6D$}
	
	Working now in a generally covariant theory where 
	spacetime is $5+1$ dimensional, the relativistic object we would like
	to consider is a $3$-brane. Furthermore, we would like to make the
	whole model \textit{conformally invariant}. This can be obtained
	provided we insert a ``composite'' scalar density $\Phi$ in place of
	$\sqrt{-g_{(5+1)}}$ in the first  term in (\ref{sact}) plus a similar
	contribution in the curvature term in the action

	\begin{eqnarray}
	S = &&-\frac{1}{16\pi G_{(5+1)}}\int d^{5+1}x \, \Phi\, g^{{}^{AB}}\,
	R_{{}_{AB}}\left(\, \Gamma\,\right)
	+e^2\,\int d^{5+1}x \,\Phi\,
	\sqrt{-\frac{1}{2\times 4!}g_{{}_{AE}}\dots g_{{}_{DH}}\, W^{{}^{ABCD}} \,
	W^{{}^{EFGH}}\,}+\nonumber\\
	&&-\frac{1}{4!}\,\int d^{5+1}x \,\sqrt{-g_{(5+1)}}\, W^{{}^{EFGH}}
	\partial_{[\, {}_{E}}\, B_{{}_{FGH}\,]}\label{confact}\\
	&&\Phi\equiv \epsilon^{A_1\dots A_6}\epsilon_{a_1\dots a_6}
	\partial_{A_1}\phi^{a_1}\dots \partial_{A_6}\phi^{a_6}
	\end{eqnarray}
	
	where, $\phi^{a_1}$,\dots $\phi^{a_6}$ are six scalar fields 
	treated as independent degrees of freedom and we consider the
	gravitational
	action in the first order formulation, i.e. $g_{{}_{AB}}$ and 
	$\Gamma^C_{DE}$ are treated as independent variables. The connection
	$\Gamma^C_{DE}$ is torsion-free, i.e. $\Gamma^C_{DE}=\Gamma^C_{ED} $.
	Thus, $\partial_{[\, {}_{E}}\, B_{{}_{FGH}\,]}\equiv \nabla_{[\,
	{}_{E}}\, B_{{}_{FGH}\,]}$ where $\nabla_{{}_M}$ is the covariant
	derivative. In
	Eq.(\ref{confact}) $R_{{}_{AB}}\equiv R^{{}^C}{}_{{}_{ABC}}$ and
	$R^{{}^A}{}_{{}_{BCD}}=\Gamma^{{}^A}_{{}_{B\, C\ , D}} 
	-\Gamma^{{}^A}{}_{{}_{B\, D\ , C}} + \Gamma^{{}^A}_{{}_{K\,D}} 
	\Gamma^{{}^K}_{{}_{B\,C}}  - \Gamma^{{}^A}_{{}_{K\,C}} 
	\Gamma^{{}^K}_{{}_{B\,D}}$.\\
	\\
	$\Phi d^{5+1}x$ is
	a scalar as well as $\sqrt{-g_{(5+1)}}\, d^{5+1}x$ under x-coordinates
	transformation, while under scalar fields re-definitions:
	
	\begin{eqnarray}
	&&\phi^{a_j}\longrightarrow \phi^{\prime\,
	b_k}\left(\,\phi^{a_j}\,\right)\label{transf}\\
	&&\Phi\longrightarrow \Phi^\prime= J\Phi\ ,\quad 
	J\equiv \mathrm{det}\left(\,
	\frac{\partial\phi^{\prime\, a_j} }{\partial\phi^{b_k}
	}\,\right)\label{transf2}
	\end{eqnarray}

	$W^{{}^{ABCD}}=$ $3$-brane slope field, it assigns a tangent
	(hyper)plane to each spacetime point; the $W$ field is \textit{totally
	anti-symmetric} in the four indices.\\
	$B_{{}_{FGH}}=$ $3$-brane \textit{gauge potential}; the $B$ field is
	\textit{totally anti-symmetric} in the three indices. In the last term
	the invariant integration measure is written in terms of $g_{(5+1)}$,
	instead of $\Phi$ to make the action invariant under (\ref{transf}).
	One must in this case assume the following Weyl rescalings also
	
	\begin{eqnarray} 
	 && g_{{}_{A_1 A_2}}\longrightarrow J\,g_{ {}_{A_1 A_2}} \\
	 && g^{{}^{B_1 B_2}}\longrightarrow J^{-1}\,g^{{}^{B_1 B_2}} \\
	 && g_{(5+1)}\longrightarrow J^6\,g_{(5+1)} \\
	 && W^{{}^{ABCD}}\longrightarrow J^{-3}\,W^{{}^{ABCD}}\\
	 && B_{{}_{FGH}}\longrightarrow B_{{}_{FGH}}\ ,\qquad 
	 \Gamma^{{}^A}{}_{{}_{BC}}\longrightarrow \Gamma^{{}^A}{}_{{}_{BC}}
	 \end{eqnarray}
	
	Notice that this symmetry holds \textit{only} in the case the embedding
	 space in $6D$. Let us remark that if we define $W$ as a
	 ``contravariant'' object
	(upper indices) and $B$ as a covariant field (lower indices), then
	the last term in the action $S$ depends on the metric only through 
	$\sqrt{-g_{(5+1)}}$ .\\
	In a previous paper , \cite{weyldp}, a conformally invariant formulation of
	the brane alone was achieved. 
        This conformally invariant theory concerns branes only.
       In this paper we are able to formulate the brane plus gravity in a 
	conformally invariant fashion. 
This is possible if the $3$-branes are embedded in a six dimensional space. Of
course it is only after the inclusion 
of gravity into the conformal invariance that the formulation can have an impact
into the question of the cosmological constant problem.\\
	
	We can define the \textit{Dual Representation} of the theory by
	 changing variables

	\begin{equation}
	W^{{}^{ABCD}} = \frac{1}{2}\frac{\epsilon^{{}^{ABCDEF}}}
	{\sqrt{-g_{(5+1)}}}\, 
	\omega_{{}_{EF}} \label{dual1}
	\end{equation}
	
	\begin{eqnarray}
	 S = &&-\frac{1}{16\pi G_{(5+1)}}\int d^{5+1}x \, 
	\Phi\, R_{(5+1)} 
	+e^2\,\int d^{5+1}x \,\Phi\,
	\sqrt{\frac{1}{4}\, g^{{}^{AE}} g^{{}^{DH}}\, \omega_{{}_{AD}} \,
	\omega_{{}_{EH}}}+\nonumber\\
	&&-\frac{1}{6!}\int d^{5+1}x\,\epsilon^{{}^{ABCDEF}}
	\omega_{{}_{[\, AB}}\partial_{{}_C}\, B_{{}_{DEF\,]}}\label{sdual}
	\end{eqnarray}
	In the next section
	we are going to use the dual formulation, defined by (\ref{sdual}),
	to obtain classical solutions describing a new kind of brane-universe.
	With this purpose in mind, it is useful to recover
	the explicit form of $\omega$ in terms of the brane density current
	
	\begin{eqnarray}
	 \omega_{{}_{MN}}&&= \frac{1}{4!}\sqrt{-g_{(5+1)}}\,
	 \epsilon_{{}_{MNABCD}}\, W^{{}^{ABCD} }\nonumber\\
	 &&= \frac{1}{4!}\,
	 \epsilon_{{}_{MNABCD}}\, \delta^{{}^A}_\mu\,\dots \delta^{{}^D}_\sigma
	 \, J^{\mu\nu\rho\sigma} \label{dualw}
	 \end{eqnarray}
	 
	 We notice that because of the total anti-symmetry of the $\epsilon$ 
	 tensor the free indices $M$, $N$ are actually projected over 
	 transverse dimensions. Accordingly, we can write (\ref{dualw}) as
	 \begin{eqnarray}
	 \omega_{{}_{MN}}&&= \rho\left(\, \vec x_\perp\,\right)
	 \,\delta^i_{{}_M} \delta^i_{{}_N} 
	 \epsilon_{ij}\,\frac{1}{4!} \epsilon_{\mu\nu\rho\sigma}\,
	 \int d^{3+1}\sigma\,\delta^{(3+1)}\left[\, x_{||}- Y_{||}(\sigma)\,
	 \right]\, dY^\mu_{||}\wedge \dots \wedge dY^\sigma_{||}\nonumber\\
	 &&=\rho\left(\, \vec x_\perp\,\right)\,
	 \epsilon_{ij}\,\delta^i_{{}_M} \delta^i_{{}_N} 
	 \frac{1}{4!} \epsilon_{\mu\nu\rho\sigma}
	 \int d^{3+1}\sigma\,\delta^{(3+1)}\left[\, x_{||}- Y_{||}(\sigma)\,
	 \right]\, dY^\mu_{||}\wedge \dots \wedge dY^\sigma_{||}\nonumber\\
	 &&=\rho\left(\, \vec x_\perp\,\right)\,
	 \epsilon_{ij}\,\delta^i_{{}_M} \delta^i_{{}_N} 
	 \frac{1}{4!} \epsilon_{\mu\nu\rho\sigma}
	 \int d^{3+1}\sigma\,\delta^{(3+1)}\left[\, x_{||}- Y_{||}\,
	 \right]\, \epsilon^{\mu\nu\rho\sigma}\frac{\partial\left(\, Y^0_{||}\ ,
	 \dots
	 Y^3_{||}\,\right) }
	 { \partial\left(\, \sigma^0\ ,\dots  \sigma^3\,\right)}
	 \nonumber\\
	 &&=-\epsilon_{ij}\,\delta^i_{{}_M}\, \delta^i_{{}_N}\, 
	 \rho\left(\, \vec x_\perp\,\right)
	 \label{final}
	 \end{eqnarray}
	
	Eq.(\ref{final}) shows that $\omega$ has non vanishing components only
	along the transverse dimensions and equals the dual of the brane
	density.

	\section{Field Equations }
	
	We will work out the equations of motion in the dual picture first and
	afterwards we will discuss the brane interpretation of these
	solutions.\\
	To start let us notice the following facts concerning the action
	(\ref{sdual}). First it can be written in the form
	
	\begin{equation}
	 S =\int d^{5+1}x \, \Phi\,\left(\, L_G + L_m\,\right)-
	\frac{1}{6!} \int d^{5+1}x \,\epsilon^{{}^{ABCDEF}}\,
	\omega_{ {}_{[\,AB}}\partial_{{}_C}\, B_{{}_{DEF\,]}}
	\end{equation}
	
	where
	
	\begin{eqnarray}
	L_G &&=- \frac{1}{ 16\pi G_{(5+1)}}\,g^{{}^{AB}}\,
	 R_{{}_{AB}}\left(\,\Gamma\,\right)\ ,\\
	L_m &&=e^2\sqrt{\frac{1}{4}\,\omega_{{}_{AB}}\,\omega_{{}_{CD}}\, 
	g^{{}^{AC}}\, g^{{}^{BD}}}
	\end{eqnarray}

	are homogeneous of degree one in $g^{{}^{AC}}$ , that is
	
	\begin{equation}
	g^{{}^{AB}}\frac{\partial L_m}{\partial g^{{}^{AB}}}=L_m\ ,\qquad
	g^{{}^{AB}}\frac{\partial L_G}{\partial g^{{}^{AB}}}=L_G
	\end{equation}
	
	this property is intimately related to the fact that the action
	(\ref{sdual}) has the symmetry under $g^{{}^{AB}}\longrightarrow 
	J^{-1}\,
	g^{{}^{AB}}$, $\Phi\longrightarrow J\, \Phi $.\\
	The equations of motion which result from the variation of the fields 
	$\phi^a$ are
	
	\begin{equation}
	\mathbf{A}^{M}_a\,\partial_{{}_M}\,\left(\, L_G + L_m\,\right)=0
	\label{phieq}
	\end{equation}

	where
	
	\begin{equation}
	\mathbf{A}^{M}_m\equiv \epsilon^{{}^{MBCDEF}} \,\epsilon_{mbcdef}\,
	\partial_{{}_B} \,\phi^b\,\partial_{{}_C} \,\phi^c\,\partial_{{}_D}\,
	 \phi^d\,\partial_{{}_E} \,\phi^e\,\partial_{{}_F} \,\phi^f
	 \end{equation}
	
	Since $\mathrm{det}\left(\, \mathbf{A}^{M}_m\,\right)=6^{-6}\Phi^6/6!$.
	Then we have that if $\Phi\ne 0$, this means that (\ref{phieq})
	implies 
	\begin{equation}
	L_G + L_m=M=\mathrm{const}.
	\end{equation}
	
	The equation of motion obtained from the variation of $g^{{}^{AB}}$ is
	
	\begin{equation}
	-\frac{1}{16\pi\, G_{(5+1)}}\, R_{{}_{AB}}+
	\frac{\partial\,L_m}{\partial g^{{}^{AB}}}=0 \label{einsteq}
	\end{equation}

	by contracting (\ref{einsteq}) with respect to $g^{{}^{AB}}$ and using 
	the homogeneity property of $L_m$, we obtain that the constant of
	integration $M$ equals zero. Evaluating 
	$\partial L_m/\partial g^{{}^{AB}}$
	and inserting into (\ref{einsteq}) we find
	
	\begin{equation} 
	R_{{}_{AB}}=4\pi\, e^2\, G_{(5+1)}\frac{\omega_{{}_{AC}}\, 
	\omega_{{}_B}{}^{{}^C}}
	{\sqrt{
	\frac{1}{4}\,\omega_{{}_{MN}}\,\omega^{{}^{MN}}\, }} \label{}
	\end{equation}

	Eq.(\ref{einsteq}) is also consistent with the Einstein form
	
	\begin{eqnarray} 
	&& R_{{}_{AB}}-\frac{1}{2}g_{{}_{AB}}\,R= -8\pi G_{(5+1)} \, 
	T_{{}_{AB}} \label{rt}\\
	&& T_{{}_{AB}}= -2\frac{\partial L_m}{\partial g^{{}^{AB}}}
	+g_{{}_{AB}}\, L_m
	\end{eqnarray}
	
	which for $L_m$ is given by
	\begin{equation} 
	T_{{}_{AB}}=\frac{e^2}{2}\frac{ \omega_{{}_{AC}}\, 
	\omega_{ {}_B}{}^{{}^C}}
	{\sqrt{\frac{1}{4}\,\omega_{{}_{MN}}\,\omega^{{}^{MN}}}
	}-e^2\, g_{{}_{AB}}\sqrt{\frac{1}{4}\,\omega_{{}_{MN}}\,
	\omega^{{}^{MN}}}\label{tab}
	\end{equation} 
	 as one can easily check that solving from $R$ by contracting both 
	 sides
	 of (\ref{rt}) with $T_{{}_{AB}}$ given by (\ref{tab}) and then 
	 replacing $R$ into (\ref{rt}) gives (\ref{einsteq}).\\
	 Let us consider now the equation of motion for the connection
	 coefficients $\Gamma^{{}^A}{}_{{}_{BC}}$.
	 Defining 
	 
	 \begin{equation}
	 \bar g_{{}_{AB}}=\left(\, \frac{\Phi}{\sqrt{-g_{(5+1)}}}\,
	 \right)^{1/2}\, g_{{}_{AB}} \label{gconf}
	 \end{equation}
	 
	 one can verify that 
	 
	 \begin{equation}
	 \Phi\, g^{{}^{AB}}=\sqrt{-\bar g_{(5+1)}}\, \bar g^{{}^{AB}}
	 \end{equation}
	 
	 Therefore, the equation of motion for $\Gamma^{{}^A}{}_{{}_{BC}}$ is 
	 obtained by the condition that the functional
	 
	 \begin{equation}
	 I\equiv -\frac{1}{16\pi\,\pi G_{(5+1)}} \int d^{5+1}x \,
	 \sqrt{-\bar g_{(5+1)}} \,\bar g^{{}^{AB}}\,
	 R_{{}_{AB}}\left(\,\Gamma\,\right)
	 \end{equation}
	 
	 is extremized under variation of $\Gamma^{{}^A}{}_{{}_{BC}}$. This is 
	 however the
	 well known Palatini problem in General Relativity (~but where the
	 metric $\bar g_{{}_{AB}}$ enters, not the original metric 
	 $g_{{}_{AB}}$ ~).
	 Therefore $\Gamma^{{}^A}{}_{{}_{BC}}$ is the well known Christoffel 
	 symbol, but not of the metric $g_{{}_{AB}}$ rather than the metric 
	 $\bar g_{{}_{AB}}$:
	 
	 \begin{equation}
	 \Gamma^{{}^A}{}_{{}_{BC}}=\left\{\,{}_B {}^A {}_C\,\right\}
	 \vert_{\bar g}
	 \end{equation}
	 
	 Notice the interesting fact that $\bar g_{{}_{AB}}$ is conformally
	 invariant, i.e.
	 invariant under the set of transformations (\ref{transf}),
	 (\ref{transf2}).\\
	 Also, in the gauge $\Phi=\sqrt{-g_{(5+1)}}$, the metric $g_{{}_{AB}}$
	 equals the metric $\bar g_{{}_{AB}}$ so one may call this the 
	 ``Einstein 
	 gauge'', since here all non-Riemannian contributions to the connection
	 disappear. Alternatively, without need of choosing a gauge one may
	 choose to work with the conformally invariant metric $\bar g_{{}_{AB}}$
	 in terms of which the connection equals the Christoffel symbol and
	 all non-Riemannian structures disappear. Finally the equations of
	 motion obtained from the variation of the gauge fields 
	 $\omega_{{}_{AB}}$ and $B_{{}_{MNP}}$ are
	 
	 \begin{equation}
	 \Phi\,\frac{\omega^{{}^{AB}}}{\sqrt{\frac{1}{4}\omega_{{}_{MN}} 
	 \omega^{{}^{MN}}}}=
	 \frac{1}{6!}\epsilon^{{}^{ABCDEF}}\partial_{[\, {}_{C}}\, 
	 B_{{}_{DEF}\,]}
	 \label{fieq}
	 \end{equation}
	 
	 and
	 \begin{equation}
	 \epsilon^{{}^{ABCDEF}}\partial_{[\, {}_{D}}\, \omega_{{}_{EF}\,]}=0
	 \label{curl}
	 \end{equation}
	 
	 taking the divergence of (\ref{fieq}) we obtain
	 
	 \begin{equation}
	 \partial_{{}_A} \left(\,
	 \Phi\,\frac{\omega^{{}^{AB}}}{\sqrt{\frac{1}{4}\omega_{{}_{MN}}
	 \omega^{{}^{MN}}}}\,
	 \right)=0 \label{diveq}
	 \end{equation}
	 
	 Again, we are looking for solution of (\ref{diveq}) where
	 the $\omega$-field is localized over the brane and not propagating
	 into in the bulk. Thus
	 
	 
	 \begin{equation}
	 \omega_{{}_{EF}}= \frac{1}{4!}
	 \epsilon_{{}_{EF ABCD}}\, J^{{}^{ABCD}} \label{dualj}
	 \end{equation}
	
	Once (\ref{dualj}) is inserted back into Eq.(\ref{curl}) one
	gets the divergence-free condition for the density current
	\begin{equation}
	\partial_{{}_M}\, J^{{}^{MBCD}}=0
	\end{equation}
	
	which is satisfied because the $D$--brane is infinitely extended.
	As one should expect, the classical solution for $\omega$ is 
	the dual of the classical solution for $W$. 
	
	\section{Brane-world solutions in the Dual Picture}
	
	In this section we are going to consider the product spacetime
	
	\begin{equation}
	ds^2=g_{\mu\nu}(x_{||})\, dx^\mu_{||}\, dx^\nu_{||}+ 
	\gamma_{ij}\left(\, \vec x_\perp\,\right)\, dx_\perp^i\, dx_\perp^j 
	\label{lelem}
	\end{equation}
	
	where $\mu\ ,\nu=0\ , 1\ , 2\ , 3$ and $i\ ,j=4\ , 5$. Furthermore,
	we consider a slope field $W^{{}^{ABCD}}$ with non-vanishing components
	only in the first four coordinates $\left(\, 0\ ,1\ ,2\ ,3\,\right)$,
	which means we are dealing with a set of parallel branes orthogonal to
	the extra-dimensions ( more on the brane interpretation of the
	solutions in the next section ). This means that the dual field
	$\omega_{AB}$ has non-zero components in the $4\ ,5$ directions only.
	In this case, we see from eq.(\ref{ricci0}) that the Ricci curvature
	induced in the four dimension $0\ ,1\ ,2\ ,3$, is zero:
	
	\begin{equation}
	R_{\mu\nu}=0\label{ricci0}
	\end{equation}
	
	Thus, the ordinary four dimensions ( accessible to our experience )
	are not curved by this kind of matter. This is a very important
	remark, since there is no need to introduce a bare cosmological constant
	to cancel some contribution from the gauge field, 
	\textit{no type of fine  tuning},  most usual in extra dimensional 
	theories, \textit{is needed here}.\\
	The simplest 
	solution of (\ref{lelem}) is \textit{flat}, four dimensional spacetime
	
	\begin{equation}
	g_{\mu\nu}=\eta_{\mu\nu}\ .
	\end{equation}
	
	Let us analyze now the additional field equations. It is convenient
	to choose gauge $\Phi=\sqrt{-g_{(5+1)}}$, even if the conformally
	invariant metric $\bar g_{{}_{AB}}$ gives the same results.\\
	The two-dimensional metric $\gamma_{ij}$ can always be put in a
	conformally flat form, i.e. one can always choose a  coordinate
	system where
	
	\begin{equation}
	\gamma_{ij}\, dx_\perp^i\, dx_\perp^j=
	\psi\left(\, x^4\ , x^5\, \right)\left[\,
	\left(\, dx^4\,\right)^2 + \left(\, dx^5\,\right)^2\,\right]
	\label{2metric}
	\end{equation}
	
	As far as the dual slope field is concerned, its most general form
	along the extra dimension where it is non-zero,
	is dictated by its tensorial structure in two-dimensions, which is
	
	\begin{equation}
	\omega^{ij}=-\frac{\epsilon^{ij}}{\sqrt\gamma}
	\rho\left(\,  x^4\ , x^5\,\right)\ , 
	\qquad \gamma\equiv \mathrm{det}\left(\,\gamma_{ij}\,\right)
	\label{omega}
	\end{equation}
	
	 It turns out that the field equations do not determine the function
	 $\rho$ as
	 
	 \begin{equation}
	 \partial_i\left(\, \frac{\omega^{ij}\,\sqrt{\gamma}}{\sqrt{-\frac{1}{2}
	 \omega^{kl}\, \omega_{kl}}}\,\right)=0\longrightarrow 
	 \partial_i\epsilon^{ij}=0
	 \end{equation}
	 
	 which is ``trivially'' satisfied $\epsilon^{ij}$ being the totally
	 anti-symmetric symbol in two-dimensions.\\
	 The function $\rho\left(\,  x^4\ , x^5\,\right)$ acts , however, as a
	 source that determines the metric.
	 The physical source of the arbitrariness in $\rho$ can be understood 
	 by invoking  
	 the brane interpretation of the $\omega$-field. The function $\rho$ is
	 associated to the density of $3$-branes being piled in the extra
	 dimensions.  Since these branes do not exert any force one upon each
	 other they can be accumulated with an arbitrary density at each
	 extra-dimensional point. Indeed, the definition of $\rho$ by 
	 equations (\ref{bdens}) and (\ref{omega}) actually are consistent.
	 Recalling that the scalar curvature of (\ref{2metric}) is 
	 $R=-\psi^{-1}\nabla^2\psi$, we have from $R=16\pi G_{(5+1)}L_m$:
	 
	 \begin{equation}
	 -\frac{1}{\psi}\nabla^2 \psi=16\pi G_{(5+1)} \,\rho
	 \label{metrica}
	 \end{equation}
	 
	 $\rho$ is free to be taken any possible values, but once it is assigned
	 $\psi$ is determined by (\ref{metrica}). The argument can be also 
	 reversed:
	 for any $\psi$ (\ref{metrica}) gives the corresponding $\rho$. An 
	 interesting
	 case is obtained when rho consists of a constant part plus one
	 or more delta function parts. Since $R$ is a scalar a delta function
	 part can appear only in combination $\delta^{(2)}/\sqrt{\gamma}$.
	 Let us define:
	 
	 \begin{eqnarray}
	 r&&=\sqrt{  (x^4)^2 + (x^5)^2 }\label{s2}\\
	 x^4&&= r\sin\phi\label{s3}\\
	 x^5&&= r\cos\phi\label{s4}
	\end{eqnarray}
	
	which describe the metric close to $r=0$, and take $\psi=\psi(r)$, so

	\begin{equation}
	\gamma_{ij} dx_\perp^i\, dx_\perp^j=
	\psi(r)\left(\, dr^2 + r^2 d\phi^2\,\right)
	 \end{equation}
	
	Then, using the representation of the delta-function ( with integration
	measure $r d\phi dr $ )
	
	 \begin{equation}
	\delta^{(2)}\left(\, r\, \right)= \frac{1}{2\pi}\nabla^2\ln r
	\end{equation}
	
	where $\nabla^2=\frac{d^2}{dr^2}+ \frac{1}{r}\,\frac{d}{dr}$.
	Then, for
	
	\begin{equation}
	\rho=\sqrt 2\, B_0 + T\frac{ \delta^{(2)}\left(\, r\, \right)}{\psi }
	\label{rho}
	\end{equation}
	
	where $B_0$ and $T$ are constants. By inserting (\ref{rho}) into 
	(\ref{metrica}) we obtain ( similar equation was obtained in 
	Ref.\cite{18} in the context of $2+1$ gravity )
	
	\begin{equation}
	\psi=\frac{4\alpha^2 b^2}{r^2}\left[\, \left(\, \frac{r}{r_0}
	\right)^\alpha +  \left(\, \frac{r}{r_0} \right)^{-\alpha}\,\right]^{-2}
	\end{equation}
	
	where
	
	 \begin{eqnarray}
	&&\alpha \equiv 1-4 G_{(5+1)}T\\
	&& b^2\equiv \frac{\sqrt 2}{16\pi G_{(5+1)}B_0}
	\end{eqnarray}

	Such a metric can be transformed into the form

	\begin{equation}
	\gamma_{ij}\, dx_\perp^i\, dx_\perp^j=b^2\,\left(\, d\theta^2 + 
	\alpha^2\,\sin^2\theta\, d\phi^2\,\right)
	 \end{equation}
	
	where $\phi$ ranges from $0$ to $2\pi$, or, equivalently,

	\begin{equation}
	\gamma_{ij}\, dx_\perp^i \, dx_\perp^j=
	b^2\,\left(\, d\theta^2 + \sin^2\theta\,
	 d\bar\phi^2\,\right)
	 \end{equation}
	
	where $\bar\phi$ now ranges from $0$ to $2\alpha\pi< 2\pi$. A complete
	solution must contain two
	branes ( in the coordinate system (\ref{s2}), (\ref{s3}),(\ref{s4})
	we are able  to display only one pole of the sphere ), the other one 
	is at the other pole of the sphere, where in $(r\ ,\phi)$ coordinates 
	is at $r\to\infty$ ). Here the term ``branes'' means delta-functions
	contributions to $\rho$.\\
	Of course, this solution is one out of a continuum of solutions, but is
	interesting because it allows us to connect to other works on the
	subject ( see Ref.\cite{15} where similar effects are discussed ).\\
	Nevertheless, we stress the fact that the function $\rho$ is totally
	free in our model. 
	
	\section{Discussion and conclusions}
	In this paper we have discussed how the gauge formulation of branes can
	be used in the framework of ``brane world'' scenarios.\\
	The formulation of $3$-branes in a six-dimensional target spacetime
	can be made in a conformally invariant way. This is possible for
	extended objects in case the target spacetime has two more dimensions
	than the extended object itself. \\
	This conformal invariance is intimately related to fact that the branes 
	( or equivalently the associated gauge fields ) only curve the manifold
	orthogonal to the brane, the extra-dimensions. No fine tuning of a
	$6D$ cosmological constant is needed in this case. Therefore, no ``old
	cosmological constant problem'' , as Weinberg has defined it \cite{sw},
	appears.\\
	An interesting phenomenon  is that the parallel $3$-branes can be found
	with an arbitrary density for any value of 
	$\vec x_\perp=\left(\, x^4\ , x^5\,\right)$. The density 
	$\rho\left(\, \vec x_\perp\,\right)$ cannot be determined . This
	represents a large degeneracy and, therefore, a freedom in the possible
	ways the branes can be accounted in the extra dimensions.\\
	The basic feature, that the matter curves only the extra dimensions
	is related to the fact that one is able to formulate the theory in terms
	of the measure $\Phi$, since then Eq.(\ref{einsteq}) follows
	automatically. Provided we adopt such formulation  Eq.(\ref{einsteq})
	tell us that if $L_m$ depends only from $\gamma_{ij}$, then only extra
	dimensions are curved.   
	Equivalence of this theory to GR requires, however, the
	existence of conformal invariance 
	since in this case one can choose the
	gauge $\Phi=\sqrt{-g_{(5+1)}}$, where the gravitational field
	equations assume the Einstein form. Conformal invariance
	holds if the embedding space is $6D$.\\
	While a conformally invariant
	formulation of the brane alone was already achieved in \cite{weyldp}, 
	it is only after the inclusion of gravity into the conformal invariance
	that the formulation can have an impact into the question of the 
	cosmological constant problem. In this paper we are able to formulate
	the brane plus gravity in a conformally invariant fashion provided
	the $3$-brane is embedded in a six dimensional space.

	Acknowledgments. \\
	E.I. Guendelman wants to thank the Department of Theoretical
	Physics of the University of Trieste for hospitality, 
	I.N.F.N. for support and Prof. M.Pavsic for discussion.


	\end{document}